\documentclass[a4,12pt]{article}

\usepackage{bm}
\usepackage{graphicx}
\usepackage{amsmath}
\usepackage{amsfonts}

\newcommand{\ol}{\overline}

\def\tr{\mathop{\rm tr}\nolimits}
\def\Pexp{\mathop{\rm Pexp}\nolimits}

\newcommand{\CC}{\mathbb{C}}

\newcommand{\nn}{\nonumber}

\begin{document}

\begin{titlepage}
\title{
\vspace{-1.5cm}
\begin{flushright}
{\normalsize TIT/HEP-677\\ Jan 2020}
\end{flushright}
\vspace{1.5cm}
\LARGE{Schur index of the ${\cal N}=4$ $U(N)$ SYM via the AdS/CFT correspondence}}
\author{
Reona {\scshape Arai\footnote{E-mail: r.arai@th.phys.titech.ac.jp}},
Shota {\scshape Fujiwara\footnote{E-mail: s.fujiwara@th.phys.titech.ac.jp}},\\
Yosuke {\scshape Imamura\footnote{E-mail: imamura@phys.titech.ac.jp}},
and
Tatsuya {\scshape Mori\footnote{E-mail: t.mori@th.phys.titech.ac.jp}}
\\
\\
{\itshape Department of Physics, Tokyo Institute of Technology}, \\ {\itshape Tokyo 152-8551, Japan}}

\date{}
\maketitle
\thispagestyle{empty}

\begin{abstract}
We calculate the Schur index of the ${\cal N}=4$ $U(N)$ SYM with finite $N$
via the AdS/CFT correspondence as the contribution of D3-branes
wrapped on contractible cycles in $\bm{S}^5$
on some assumptions motivated by preliminary analyses.
As far as we have checked numerically it agrees with the index calculated on the gauge theory side.
In a certain limit it reproduces
the analytic result given by 
Bourdier, Drukker, and Felix.
\end{abstract}
\end{titlepage}
\section{Introduction}
The Schur index \cite{Gadde:2011uv} is a specialization of the superconformal index \cite{Kinney:2005ej}
that is defined for ${\cal N}=2$ superconformal theories.
It is a function of a universal fugacity $q$,
and if the theory has a flavor symmetry we can also introduce additional flavor fugacities.
For a Lagrangian theory we can calculate it by the localization formula,
which gives the index as a matrix integral.
In some cases we can use other methods.
The IR formula \cite{Cordova:2015nma}
enables us to calculate it
from the BPS spectrum in the Coulomb branch.
For a class S theory it is given as a correlation function on a Riemann surface
\cite{Gadde:2009kb,Gadde:2011ik}.
It is also known that for an arbitrary ${\cal N}=2$ superconformal theory
there is a corresponding chiral algebra
and the Schur index is given as the vacuum character of the chiral algebra \cite{Beem:2013sza}.
By using these different methods
complementarily we can obtain non-perturbative information of the theory.
In this work we propose another method to calculate the Schur index
of the ${\cal N}=4$ $U(N)$ SYM based on the AdS/CFT correspondence \cite{Maldacena:1997re}.

The ${\cal N}=4$ theory, regarded as a special ${\cal N}=2$ theory, has the flavor symmetry $SU(2)_F\subset SU(4)_R$
and we introduce the flavor fugacity $u$.
The Schur index is defined by
\begin{align}
{\cal I}(q,u)=\tr_{\rm BPS}(e^{2\pi i(J+\ol J)}q^{H+J+\ol J}u^{R_x-R_y}),
\label{schurdef}
\end{align}
where the trace is taken over states saturating certain bounds.
See \cite{Arai:2019xmp} for our conventions.
The localization formula is
\begin{align}
{\cal I}_{U(N)}(q,u)=\int d\mu_N\Pexp(i_V(q,u)\chi_N(z_a)),
\label{iuneq}
\end{align}
where $d\mu_N$ is the $U(N)$ Haar measure of the integral over gauge fugacities $z_a$ ($a=1,\ldots,N$)
and $\chi_N(z_a)$
is the character of the $U(N)$ adjoint representation:
\begin{align}
d\mu_N=\frac{1}{N!}\prod_{a=1}^N\frac{dz_a}{2\pi iz_a}\prod_{a\neq b}\left(1-\frac{z_a}{z_b}\right),\quad
\chi_N(z_a)=\sum_{a,b=1}^N\frac{z_a}{z_b}.
\end{align}
We define the plethystic exponential $\Pexp$ by
\begin{align}
\Pexp f=\prod_i \frac{1}{(1-f_i)^{c_i}},
\label{pexpdef}
\end{align}
for a function $f$ with the series expansion
$f=\sum_ic_if_i$,
where $c_i$ are numerical coefficients and $f_i$ are monomials made of the fugacities.
The letter index $i_V$ is
\begin{align}
i_V(q,u)=\frac{q(u+\frac{1}{u})-2q^2}{1-q^2}.
\label{fbdr}
\end{align}
For the special case with $u=1$
Bourdier, Drukker, and Felix
analytically carried out the integral in (\ref{iuneq}) and obtain \cite{Bourdier:2015wda}
\begin{align}
\left.\frac{{\cal I}_{U(N)}}{{\cal I}_{U(\infty)}}\right|_{u\rightarrow 1}
=
\sum_{n=0}^\infty{\cal I}_n^{\rm BDF},\quad
{\cal I}_n^{\rm BDF}
=
(-1)^n
({}_{N+n}C_N+{}_{N+n-1}C_N)q^{nN+n^2},
\label{BDFformula}
\end{align}
where $_nC_k=\frac{n !}{k! (n-k)!}$ is the binomial coefficient.

The purpose of this paper is to reproduce
(\ref{iuneq}) and (\ref{BDFformula}) for finite $N$ on the AdS side.
In the large $N$ limit we can analytically evaluate the integral
by the saddle point analysis \cite{Kinney:2005ej}, and obtain
\begin{align}
{\cal I}_{U(\infty)}(q,u)
=\Pexp\left(\frac{uq}{1-uq}+\frac{u^{-1}q}{1-u^{-1}q}-\frac{q^2}{1-q^2}\right).
\label{iinf}
\end{align}
On the AdS side this is reproduced as the index of Kaluza-Klein modes of the supergravity multiplet
in the dual spacetime $AdS_5\times\bm{S}^5$.
If $N$ is finite, as the parameter relation $N=L^4T_{\rm D3}$
including the AdS radius $L$ and the D3-brane tension $T_{\rm D3}$ implies,
we should take account of D3-branes extended in $AdS_5\times\bm{S}^5$,
and we can guess that the ratio ${\cal I}_{U(N)}/{\cal I}_{U(\infty)}$ expresses the contribution of D3-branes.
What brane configurations should we take into account to calculate the index?
In the case of BPS partition function it is possible to reproduce the exact result
by the geometric quantization of $1/8$ BPS brane configurations \cite{Biswas:2006tj}.
An $1/8$ BPS configuration
is given as the intersection of a holomorphic surface $h(X,Y,Z)=0$ and $\bm{S}^5=\{(X,Y,Z)||X|^2+|Y|^2+|Z|^2=1\}$ \cite{Mikhailov:2000ya}.
In the calculation of the superconformal index in \cite{Arai:2019xmp}
such configurations were treated as excitations of ``rigid D3-branes.''
A rigid D3-brane here means a D3-brane wrapped on a large $\bm{S}^3$ in $\bm{S}^5$
given by the linear equation $aX+bY+cZ=0$.
The collective motion of a rigid D3-brane is described by the moduli space $\CC\bm{P}^2$
with the projective coordinates $(a,b,c)$.
Corresponding to the fact that $\CC\bm{P}^2$ is covered by three coordinate patches
we can treat all rigid brane configurations and excitations of them
as excitations of three specific brane configurations: $X=0$, $Y=0$, and $Z=0$.
In the case of Schur index
only two configurations $X=0$ and $Y=0$ give non-trivial contributions,
and a part of the finite $N$ correction of the Schur index
was correctly reproduced as the contribution from a single D3-brane wrapped on
$X=0$ and $Y=0$.
See \cite{Arai:2019xmp} for more details.
Although we do not have any proof this fact seems to suggest that some localization mechanism works
for D3-brane configurations.
By assuming this mechanism keeps working for multiple-brane configurations we propose the relation
\begin{align}
\frac{{\cal I}_{U(N)}(q,u)}{{\cal I}_{U(\infty)}(q,u)}=\sum_{n_1,n_2=0}^\infty{\cal I}_{(n_1,n_2)}(q,u;N),
\label{id3}
\end{align}
where ${\cal I}_{(n_1,n_2)}$ is the contribution from
the configuration with  $n_1$ D3-branes
wrapped on $X=0$ and $n_2$ D3-branes wrapped on $Y=0$.

Classically, the energy of the brane system is $(n_1+n_2)N$ in the unit of $L^{-1}$,
and expected to give ${\cal O}(q^{(n_1+n_2)N})$ terms in the index.
By comparing this with (\ref{BDFformula})
it is natural to identify ${\cal I}_n^{\rm BDF}$
with the contribution of brane systems with $n_1+n_2=n$.
Namely
\begin{align}
{\cal I}_n^{\rm BDF}(q;N)=\lim_{u\rightarrow 1}\sum_{k=0}^n {\cal I}_{(n-k,k)}(q,u;N).
\label{maineq}
\end{align}
In the following we calculate ${\cal I}_{(n-k,k)}$ and numerically confirm
that (\ref{id3}) and (\ref{maineq}) indeed hold.

\section{Gauge theory on wrapped branes}
The brane system giving ${\cal I}_{(n-k,k)}$ consists of 
$n-k$ D3-branes wrapped on $X=0$ and $k$ D3-branes
wrapped on $Y=0$.
These two $3$-cycles intersect in $\bm{S}^5$ along $\bm{S}^1$,
and a bi-fundamental hypermultiplet arises on the intersection.
Namely, the theory realized on the brane system is
the $U(n-k)\times U(k)$ gauge theory with a bi-fundamental hypermultiplet.
The index is
\begin{align}
{\cal I}_{(n-k,k)}=(uq)^{(n-k)N}(u^{-1}q)^{kN}\int d\mu_{n-k}\int d\mu'_k\Pexp f_{\rm tot},
\label{intzz}
\end{align}
where the prefactors $(uq)^{(n-k)N}$ and $(u^{-1}q)^{kN}$ are the classical contributions
of the D3-branes wrapped on the two cycles \cite{Arai:2019xmp}.
The total letter index $f_{\rm tot}$ is
\begin{align}
f_{\rm tot}=f_V(q,u)\chi_{n-k}^{\rm adj}(z)
+f_H(q,u)\chi_{n-k,k}^{\rm bf}(z,z')
+f_V(q,u^{-1})\chi_k^{\rm adj}(z'),
\end{align}
where $f_V(q,u)$ and $f_H(q,u)$ are the letter indices
for a vector multiplet on $X=0$ and a half hypermultiplet on the intersection, respectively.
The letter index of the vector multiplet on $Y=0$ is obtained from that for $X=0$ by the
$SU(2)_F$ Weyl reflection $u\rightarrow u^{-1}$.
$\chi^{\rm bf}_{n-k,k}$ is the character of the bi-fundamental representation:
\begin{align}
\chi^{\rm bf}_{n-k,k}(z,z')=\sum_{a=1}^{n-k}\sum_{b=1}^k\left(\frac{z_a}{z'_b}+\frac{z_b'}{z_a}\right).
\end{align}
We can easily determine the BPS spectrum of the hypermultiplet
by using the supersymmetry algebra,
and we obtain the letter index
\begin{align}
f_H=\frac{1}{q}-q.
\end{align}
Fortunately, the plethystic exponential of $f_H\chi^{\rm bf}_{n-k,k}$
is quite simple:
\begin{align}
\Pexp(f_H(q)\chi^{\rm bf}_{n-k,k}(z,z'))=q^{2(n-k)k}.
\end{align}
Because this is independent of the gauge fugacities
the integral in (\ref{intzz}) is factorized into the $U(n-k)$ part and the $U(k)$ part,
and ${\cal I}_{(n-k,k)}$ is given by
\begin{align}
{\cal I}_{(n-k,k)}=(uq)^{(n-k)N}F_{n-k}(q,u)\cdot q^{2(n-k)k} \cdot (u^{-1}q)^{kN} F_k(q,u^{-1}).
\label{inkk}
\end{align}
$F_n(q,u)$ is the index of the $U(n)$ gauge theory
realized on $X=0$, which is given by
\begin{align}
F_n(q,u)=\int d\mu_n \Pexp (f_V(q,u) \chi_n(z_a)).
\label{integral}
\end{align}

As is pointed out in \cite{Arai:2019xmp}
the letter index
$f_V$ is obtained from $i_V$ in (\ref{fbdr})
by the variable change
\begin{align}
f_V(q,u)=i_V(q^{\frac{1}{2}}u^{-\frac{1}{2}},q^{-\frac{3}{2}}u^{-\frac{1}{2}})
=\frac{\frac{1}{uq}-\frac{2}{u}q+q^2}{1-\frac{1}{u}q}.
\label{varchange}
\end{align}
Correspondingly,
$F_n$ is related to ${\cal I}_{U(n)}$ by
$F_n(q,u)={\cal I}_{U(n)}(q^{\frac{1}{2}}u^{-\frac{1}{2}},q^{-\frac{3}{2}}u^{-\frac{1}{2}})$.
Unfortunately,
we cannot directly obtain $F_n$ in the form of $q$-expansion
by using this relation
as far as ${\cal I}_{U(n)}$ is also given as the $q$-expansion.
We need to calculate $F_n$ separately by
performing the integral in (\ref{integral}).
When we calculate ${\cal I}_{U(N)}$ by (\ref{iuneq}) we
usually assume $|q|<1$ and pick up poles in the unit circle on $z_a$-planes.
For the integral in (\ref{integral}) we have to pick up
the corresponding poles obtained by the variable change, which are not always in the unit circle.
We show the first few terms of $F_n$ ($n\leq4$):
\begin{align}
F_0(q,u)
&=1,\nonumber\\
F_1(q,u)
&=
\frac{u^3}{1-u^2}q
+(1-u^2)q^2
+\left(\frac{1}{u^3}-u^3\right)q^3
+\left(\frac{1}{u^6}-\frac{1}{u^2}+1-u^4\right)q^4+\cdots,\nonumber\\
F_2(q,u)
&=\frac{u^{10}(2-u^2)}{(1-u^2)(1-u^4)}q^4
+u^5(2-u^4)q^5
+(2+2u^6-u^{12})q^6
+\cdots,\nonumber\\
F_3(q,u)
&=\frac{u^{21}(5-3u^2-3u^4+2u^6)}{(1-u^2)(1-u^4)(1-u^6)}q^9
+\frac{u^{14}(5-3u^4-3u^6+2u^{10})}{1-u^4}q^{10}
+\cdots,\nonumber\\
F_4(q,u)
&=\frac{u^{36}(14-9u^2-10u^4-2u^6+6u^8+7u^{10}-5u^{12})}{(1-u^2)(1-u^4)(1-u^6)(1-u^8)}q^{16}+\cdots
.\label{f23}
\end{align}
We point out that the leading terms of $F_n$
are given by
$F_n(q,u)=q^{n^2}u^{2n^2}G_n(u)+{\cal O}(q^{n^2+1})$ with the functions $G_n$
generated by
\begin{align}
\sum_{n=0}^\infty G_n(u)t^n=\exp\left(\sum_{p=1}^\infty\frac{(2p-1)!}{(p!)^2}\frac{t^p}{u^{-p}-u^p}\right).
\end{align}

\section{Comparisons}
Let us introduce the notation $A^{(\leq m)}$ to mean
the $q$-expansion of $A$ up to the $q^m$ term.
$F_n$ contributes to the $q^{nN+n^2}$ or higher order terms in (\ref{inkk}).
To do a non-trivial check for the leading term of $F_4$ in (\ref{f23}) for $U(1)$ theory
we need to calculate the both hand sides of (\ref{inkk})
up to $q^{20}$ terms.
For this purpose we calculated $F_n^{(\leq 20-n)}$ ($n\leq 4$),
and by substituting them into the conjectural relation (\ref{id3})
we obtained
$({\cal I}_{U(N)}/{\cal I}_{U(\infty)})^{(\leq 19+N)}$ for $N=1,2,3,4$.
We have found the complete agreement with
the results obtained from (\ref{iuneq}).
(See appendix for the first few terms of ${\cal I}_{U(N)}/{\cal I}_{U(\infty)}$
calculated by (\ref{iuneq}) for small $N$.)
For $N=0$, the ``$U(0)$'' gauge theory is the trivial theory with no excitation,
and the index is ${\cal I}_{U(0)}=1$.
Although the physical interpretation on the gravity side is not clear
we have found that (\ref{id3}) with $N=0$ correctly gives
$(1/{\cal I}_{U(\infty)})^{(\leq19)}$.
We also found that
the right hand side of (\ref{id3}) vanishes for $N=-1$.
Namely, (\ref{id3}) formally gives ${\cal I}_{U(-1)}=0$.

We also confirmed (\ref{maineq}) by taking the $u\rightarrow 1$ limit.
Note that the limit must be taken after the summation with respect to $k$
because functions $F_n$ have poles at $u=1$.
If we sum up ${\cal I}_{(n-k,k)}$ over $k=0,\ldots,n$ the poles at each order
cancel and we obtain the
leading term (\ref{BDFformula}) as well as vanishing sub-leading terms.
We have confirmed that (\ref{maineq})
 correctly reproduces
$({\cal I}_n^{\rm BDF})^{(\leq n(N-1)+20)}$ for $n=1,2,3,4$ and arbitrary $N$.

\section{Discussions}

We proposed the relations (\ref{id3}) and (\ref{maineq}) for
the Schur index of the ${\cal N}=4$ $U(N)$ SYM,
and numerically confirmed that they correctly reproduce the
results obtained on the gauge theory side.

Our calculation was based on some assumptions.
We assumed the localization of the path integral to the special configurations
consisting of branes wrapped on the two specific cycles $X=0$ and $Y=0$.
We also assumed that quantum gravity corrections do not spoil our calculation.

There are many directions of extension.
There seems no essential difficulty to generalize our analysis to the superconformal index.
It would be also possible to apply our method to other examples of AdS/CFT.
An analytic formula of the Schur index for a class of ${\cal N}=2$ was obtained in \cite{Bourdier:2015sga}
and it may be possible to reproduce it by the D3-brane analysis.
There were some analysis
of single-brane configurations 
for S-fold theories \cite{Arai:2019xmp},
orbifold theories \cite{Arai:2019wgv},
and toric gauge theories \cite{Arai:2019aou}.
It would be interesting to extend these results to multiple-brane configurations.

\section*{Acknowledgements}
We would like to thank Daisuke Yokoyama for valuable discussions and comments.

\appendix
\section{Results on the gauge theory side}
In this appendix we show the explicit form of $q$-expansion of
${\cal I}_{U(\infty)}$ and ${\cal I}_{U(N)}/{\cal I}_{U(\infty)}$
calculated on the gauge theory side.

The $q$-expansion of the index in the large $N$ limit (\ref{iinf}) is
\begin{align}
&{\cal I}_{U(\infty)}(q,u)
\nonumber\\
&=
1+ \chi _1q+ \left(2 \chi _2-2\right)q^2+ \left(3 \chi _3-2 \chi _1\right)q^3+ \left(-4 \chi _2+5 \chi _4+1\right)q^4
\nn\\&\quad   
+
   \left(\chi _1-5 \chi _3+7 \chi _5\right)q^5+ \left(3 \chi _2-9 \chi _4+11 \chi _6-1\right)q^6
\nn\\&\quad      
   + \left(\chi _1+2 \chi
   _3-11 \chi _5+15 \chi _7\right)q^7+ \left(-2 \chi _2+6 \chi _4-18 \chi _6+22 \chi _8+4\right)q^8
\nn\\&\quad      
   + \left(2 \chi _3+5
   \chi _5-23 \chi _7+30 \chi _9\right)q^9+\cdots,
\label{iuinftyu}
\end{align}
where
$\chi_n=(u^{n+1}-u^{-(n+1)})/(u-u^{-1})$
is the $SU(2)$ character.
The inverse of 
(\ref{iuinftyu}) is
\begin{align}
&1/{\cal I}_{U(\infty)}(q,u)
\nonumber\\
&=
1-\chi _1q + \left(3-\chi _2\right)q^2+ \left(5-\chi _2\right)q^4+ \left(-\chi _1-\chi _3+\chi _5\right)q^5
\nn\\&\quad   
+
   \left(-\chi _2-\chi _4+8\right)q^6+ \left(\chi _7-2 \chi _3\right)q^7+ \left(-2 \chi _2-\chi _6+13\right)q^8
\nn\\&\quad      
   +
   \left(-\chi _1-2 \chi _3+\chi _7\right)q^9+ \left(-3 \chi _2-\chi _4-\chi _6+21\right)q^{10}
\nn\\&\quad      
   + \left(2 \chi _7-4 \chi
   _3\right)q^{11}+ \left(-3 \chi _2-\chi _4-2 \chi _6+\chi _{10}-\chi _{12}+30\right)q^{12}+\cdots.
\end{align}
The ratio of ${\cal I}_{U(N)}$ calculated
by using the localization formula (\ref{iuneq})
and the large $N$ limit (\ref{iuinftyu}) is
given for small $N$ as follows.
\begin{align}
&{\cal I}_{U(1)}(q,u)/{\cal I}_{U(\infty)}(q,u)
\nonumber\\
&=
1-\chi _2q^2 + \left(2 \chi _1-\chi _3\right)q^3+\left(2 \chi _2-\chi _4-1\right)q^4 +\left(2 \chi _2-1\right)q^6 
\nn\\&\quad   
+\left(\chi _3-2 \chi _5+\chi _7\right)q^7+ \left(-\chi _2-\chi _6+\chi _8+1\right)q^8
\nn\\&\quad      
   +\left(-2 \chi _1+4 \chi _3-\chi
   _5-2 \chi _7+\chi _9\right)q^9 +\left(\chi _2-\chi _6-\chi _8+\chi _{10}+2\right)q^{10} 
\nn\\&\quad      
   + \left(\chi _5-\chi _7-\chi
   _9+\chi _{11}\right)q^{11}+\left(2 \chi _4-\chi _6-\chi _8-1\right)q^{12} +\cdots,
\end{align}
\begin{align}
&{\cal I}_{U(2)}(q,u)/{\cal I}_{U(\infty)}(q,u)
\nonumber\\
&=
1-\chi _3q^3 + \left(2 \chi _2-\chi _4-1\right)q^4+ \left(\chi _1+\chi _3-\chi _5\right)q^5+ \left(2 \chi _4-\chi
   _6-3\right)q^6
\nn\\&\quad         
   +q^8 \left(\chi _2+\chi _4+1\right)q^8+ \left(-3 \chi _1+\chi _3-\chi _7+\chi _9\right)q^9
\nn\\&\quad         
   +\left(-2 \chi
   _2+\chi _4+\chi _6-2 \chi _8+\chi _{10}+1\right)q^{10} 
\nn\\&\quad         
   + \left(\chi _1+2 \chi _3-\chi _5-\chi _7-2 \chi _9+2 \chi
   _{11}\right)q^{11}+\cdots,
\end{align}
\begin{align}
&{\cal I}_{U(3)}(q,u)/{\cal I}_{U(\infty)}(q,u)
\nonumber\\
&=
1- \chi _4q^4+ \left(-\chi _1+2 \chi _3-\chi _5\right)q^5+ \left(\chi _4-\chi _6+2\right)q^6
\nn\\&\quad         
+ \left(-3 \chi _1+2 \chi
   _3+\chi _5-\chi _7\right)q^7+ \left(-\chi _4+2 \chi _6-\chi _8\right)q^8
\nn\\&\quad            
   + \left(3 \chi _2-\chi _4+2 \chi
   _6-4\right)q^{10}+ \left(-2 \chi _1+\chi _3+\chi _5-\chi _7-\chi _9+\chi _{11}\right)q^{11}
\nn\\&\quad            
   + \left(-\chi _2-3 \chi _4+2
   \chi _6-\chi _{10}+\chi _{12}+2\right)q^{12}+\cdots  ,
\end{align}
\begin{align}
&{\cal I}_{U(4)}(q,u)/{\cal I}_{U(\infty)}(q,u)
\nonumber\\
&=
1- \chi _5q^5+ \left(-\chi _2+2 \chi _4-\chi _6\right)q^6+ \left(\chi _1+\chi _5-\chi _7\right)q^7
\nn\\&\quad
+\left(-\chi _2+\chi
   _4+\chi _6-\chi _8\right)q^8
+ \left(-\chi _3+\chi _5+\chi _7-\chi _9\right)q^9
\nn\\&\quad   
   + \left(-2 \chi _2+\chi _4-\chi _6+2
   \chi _8-\chi _{10}+1\right)q^{10}
\nn\\&\quad   
   + \left(-2 \chi _2+4 \chi _4-2 \chi _6+2 \chi _8+2\right)q^{12}+\cdots   .
\end{align}


\begin{thebibliography}{99}
\bibitem{Gadde:2011uv} 
  A.~Gadde, L.~Rastelli, S.~S.~Razamat and W.~Yan,
  Commun.\ Math.\ Phys.\  {\bf 319}, 147 (2013)
  doi:10.1007/s00220-012-1607-8
  [arXiv:1110.3740 [hep-th]].
\bibitem{Kinney:2005ej} 
  J.~Kinney, J.~M.~Maldacena, S.~Minwalla and S.~Raju,
  Commun.\ Math.\ Phys.\  {\bf 275}, 209 (2007)
  doi:10.1007/s00220-007-0258-7
  [hep-th/0510251].

\bibitem{Cordova:2015nma} 
  C.~Cordova and S.~H.~Shao,
  JHEP {\bf 1601}, 040 (2016)
  doi:10.1007/JHEP01(2016)040
  [arXiv:1506.00265 [hep-th]].

\bibitem{Gadde:2009kb} 
  A.~Gadde, E.~Pomoni, L.~Rastelli and S.~S.~Razamat,
  JHEP {\bf 1003}, 032 (2010)
  doi:10.1007/JHEP03(2010)032
  [arXiv:0910.2225 [hep-th]].

\bibitem{Gadde:2011ik} 
  A.~Gadde, L.~Rastelli, S.~S.~Razamat and W.~Yan,
  Phys.\ Rev.\ Lett.\  {\bf 106}, 241602 (2011)
  doi:10.1103/PhysRevLett.106.241602
  [arXiv:1104.3850 [hep-th]].
\bibitem{Beem:2013sza} 
  C.~Beem, M.~Lemos, P.~Liendo, W.~Peelaers, L.~Rastelli and B.~C.~van Rees,
  Commun.\ Math.\ Phys.\  {\bf 336}, no. 3, 1359 (2015)
  doi:10.1007/s00220-014-2272-x
  [arXiv:1312.5344 [hep-th]].

\bibitem{Maldacena:1997re}
  J.~M.~Maldacena,
  Int.\ J.\ Theor.\ Phys.\  {\bf 38}, 1113 (1999)
  [Adv.\ Theor.\ Math.\ Phys.\  {\bf 2}, 231 (1998)]
  doi:10.1023/A:1026654312961, 10.4310/ATMP.1998.v2.n2.a1
  [hep-th/9711200].

\bibitem{Arai:2019xmp} 
  R.~Arai and Y.~Imamura,
  PTEP {\bf 2019}, no. 8, 083B04 (2019)
  doi:10.1093/ptep/ptz088
  [arXiv:1904.09776 [hep-th]].


\bibitem{Bourdier:2015wda} 
  J.~Bourdier, N.~Drukker and J.~Felix,
  JHEP {\bf 1511}, 210 (2015)
  doi:10.1007/JHEP11(2015)210
  [arXiv:1507.08659 [hep-th]].

\bibitem{Biswas:2006tj} 
  I.~Biswas, D.~Gaiotto, S.~Lahiri and S.~Minwalla,
  JHEP {\bf 0712}, 006 (2007)
  doi:10.1088/1126-6708/2007/12/006
  [hep-th/0606087].

\bibitem{Mikhailov:2000ya} 
  A.~Mikhailov,
  JHEP {\bf 0011}, 027 (2000)
  doi:10.1088/1126-6708/2000/11/027
  [hep-th/0010206].






\bibitem{Bourdier:2015sga}
  J.~Bourdier, N.~Drukker and J.~Felix,
  JHEP {\bf 1601}, 167 (2016)
  doi:10.1007/JHEP01(2016)167
  [arXiv:1510.07041 [hep-th]].

\bibitem{Arai:2019wgv} 
  R.~Arai, S.~Fujiwara, Y.~Imamura and T.~Mori,
  JHEP {\bf 1910}, 243 (2019)
  doi:10.1007/JHEP10(2019)243
  [arXiv:1907.05660 [hep-th]].

\bibitem{Arai:2019aou} 
  R.~Arai, S.~Fujiwara, Y.~Imamura and T.~Mori,
  arXiv:1911.10794 [hep-th].

\end{thebibliography}
\end{document}